# LOSSLESS DIGITAL IMAGE COMPRESSION METHOD FOR BITMAP IMAGES

Dr T. Meyyappan[1], SM.Thamarai[2] and N.M.Jeya Nachiaban[3]

[1,2] Department of Computer Science and Engineering,
Alagappa University, Karaikudi – 630 003, India.
`meyslotus@yahoo.com`, `lotusmeys@yahoo.com`

[3] Department of Computer Science and Engineering,
Thiagarajar College of Engineering Madurai-9., India.
`nmjeyan2009@tce.edu`

## ABSTRACT

*In this research paper, the authors propose a new approach to digital image compression using crack coding This method starts with the original image and develop crack codes in a recursive manner, marking the pixels visited earlier and expanding the entropy in four directions. The proposed method is experimented with sample bitmap images and results are tabulated. The method is implemented in uni-processor machine using C language source code.*

## 1. INTRODUCTION

Now a days network plays an important role in our life. It is hard to pass a day without sharing information with others. Transmission of images is also the need of the day. Transmission of images in their original form increases the time spent in network and we need to increases the bandwidth for fast transmission. On the other hand, compressed images which can be restored at the receiving end can very much reduce network overheads.

Compression of images is concerned with storing them in a form that does not take up so much space as original. Elementary compression can be illustrated by comparing the work of artists with that of computer aided designers.

This is not how artists would work. With an electronic paint brush they might wish to shade an area with a particular ink, gradually increasing the amount of shading by using the paint brush over some part of the area again and again. The file would have to be a full memory map of all the pixels in the original piece of work. It could not simply contain the (x,y) coordinates of areas and their relationships. Compression of images[9] is concerned with storing them in a form that does not take up so much space as original.

Different data structures are best for each type of image for minimum storage requirements. Minimum storage is not a useful end in itself and compressing images to minimum storage levels is itself a time-costly exercise.However, if images are to be held on or transferred between machines ,minimum storage reduces hardware costs.Image compression addresses the problem of reducing the amount of data[2] required to represent a digital image. The underlying basis of the reduction process is the removal of redundant data.

## 2. EXISTING METHODS

The four different approaches[3],[5] to compression are Statistical Compression,Spatial compression, Quantizing compression, Fractal compression. In spatial approach, image coding is based on the spatial relationship between pixels of predictably similar types. The method proposed in this paper employs spatial approach for compression.

Run-length encoding (RLE) is a very simple form of data compression in which runs of data (that is, sequences in which the same data value occurs in many consecutive data elements) are stored as a single data value and count, rather than as the original run. This is most useful on data that contains many such runs: for example, simple graphic images[8] such as icons, line drawings, and animationsHuffman coding removes coding redundancy. Huffman's procedure creates the optimal code for a set of symbols and probabilities subject to the constraint that the symbols be coded one at a time. After the code has been created, coding and/or decoding is accomplished in the simple look-up table . When large number of symbols is to be coded, the construction of the optimal binary Huffman code is a difficult task.

Arithmetic coding (AC)[4] is a special kind of entropy coding. Arithmetic coding is a form of variable-length entropy encoding used in lossless data compression. Arithmetic coding differs from other forms of entropy encoding such as Huffman coding in that rather than separating the input into component symbols and replacing each with a code, arithmetic coding encodes the entire message into a single number.

## 3. IMAGE MODEL

A digitized image is described by an N x M matrix of pixel values are nonnegative scalars, that indicate the light intensity of the picture element at (i,j) represented by the pixel.

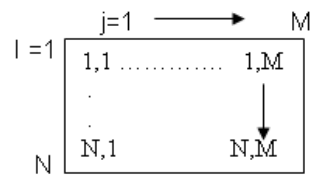

**Figure 1 Image Model**

### 3.1 Connectivity

In many circumstances it is important to know whether two pixels are connected to each other, and there are two major rules[7] for deciding this. Consider a pixel called P, at row i and column j of an image; looking at a small region centered about this pixel, we can label the neighboring pixels with integers. Connectivity is illustrated below:

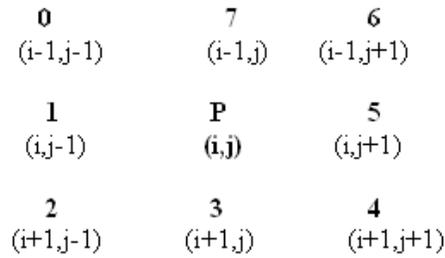

Two pixels are 4-adjacent if they are horizontal or vertical neighbors. The 4-adjacent pixels are said to be connected if they have the same pixel value.

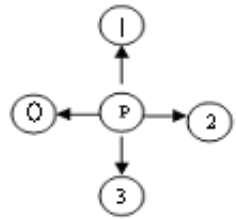

**Figure 2  4-Connected Pixels**

### 3.2 Contour Tracing with  Crack Coding

Contour coding[4] has the effect of reducing the areas of pixels of the same grey levels to a set of contours that bound those areas.  If the areas of same grey level are large with a simple edge, then the compression rate can be very good.  In practice, it is best to make all contours circular[4], so that they return to the originating pixel - if necessary along the path that they have already traversed - and to identify the grey level that they lie on and enclose.  8-connected contour is known as chain coding and 4-connected contour is known as crack coding.  In this paper, authors used crack coding and grey level of each contour is saved along with the contour.

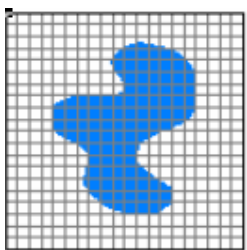 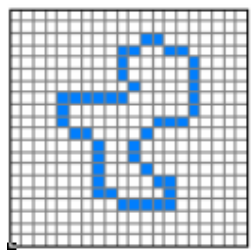

**Figure 3(a) Original Image**     **Figure3(b) 4-Connected Contour**

## 4.  BITMAP-FILE STRUCTURE

Each bitmap file contains a bitmap-file header, a bitmap-information header, a color table, and an array of bytes that defines the bitmap bits. The file has the following form:

BITMAPFILEHEADER    bmfh;

BITMAPINFOHEADER   bmih;

RGBQUAD                    aColors[];

BYTE                            aBitmapBits[];

### 4.1 Bitmap file Header

The bitmap-file header (bmfh) contains information about the type, size, and layout of a device-independent bitmap file[6]. The header is defined as a BITMAPFILEHEADER structure. Its contents and their meaning or shown in the table given below:

Table 1 Bitmap Header Structure

| START | SIZE | NAME | STANDARD VALUE | PURPOSE |
|---|---|---|---|---|
| 1 | 2 | bfType | 19778 | must always be set to 'BM' for bitmap file |
| 3 | 4 | bfSize | ?? | specifies the size of files in bytes |
| 7 | 2 | bfReaseved1 | 0 | must always be set to zero |
| 9 | 2 | bfReserved2 | 0 | must always be set to zero |
| 11 | 4 | bfOffBits | 1078 | specifies the offset from the beginning of the file to the bitmap data |

### 4.2 Bitmap Info Header

The bitmap-information header (bmih), defined as a BITMAPINFOHEADER structure, specifies the dimensions, compression type, and color format for the bitmap[6]. Its contents and their meaning are shown in the table given below.

Table 2 Bitmap File Header Structure

| START | SIZE | NAME | STANDARD VALUE | PURPOSE |
|---|---|---|---|---|
| 15 | 4 | biSize | 40 | specifies the bitmap size |
| 19 | 4 | biwidth | 100 | specifies the image width |

| 23 | 4 | biHeight | 100 | specifies the image height |
|---|---|---|---|---|
| 27 | 2 | biPlanes | 1 | specifies the number of planes |
| 29 | 2 | biBitcount | 8 | specifies the number of bits per pixel |
| 31 | 4 | biCompression | 0 | secifies the type of compression |
| 35 | 4 | bisizeImage | 0 | specifies the size of image data |
| 39 | 4 | biXPelsPerMeter | 0 | specifies the pixels in X direction |
| 43 | 4 | biXPelsPerMeter | 0 | specifies the pixels in Y direction |
| 47 | 4 | biClrUsed | 0 | specifies the number of bitmap colors |
| 51 | 4 | biClrImportant | 0 | specifies the number of colors ,important for bitmap |

### 4.3 Color Table

The color table, defined as an array of RGBQUAD structures, contains as many elements as there are colors in the bitmap. The color table is not present for bitmaps with 24 color bits because each pixel is represented by 24-bit red-green-blue (RGB) values in the actual bitmap data area.

The colors in the table should appear in order of importance. This helps a display driver render a bitmap on a device that cannot display as many colors as here are in the bitmap. If the DIB is in Windows version 3.0 or later format, the driver can use the biClrImportant member of the BITMAPINFOHEADER structure to determine which colors are important.The BITMAPINFO structure can be used to represent a combined bitmap-information header and color table.

### 4.4 Bitmap Pixels

The bitmap bits, immediately following the color table, consist of an array of BYTE values representing consecutive rows, or "scan lines," of the bitmap. Each scan line consists of consecutive bytes representing the pixels in the scan line, in left-to-right order. The number of bytes representing a scan line depends on the color format and the width, in pixels, of the bitmap. If necessary, a scan line must be zero-padded to end on a 32-bit boundary. However, segment boundaries can appear anywhere in the bitmap[6]. The scan lines in the bitmap are stored from bottom up. This means that the first byte in the array represents the pixels in the lower-left corner of the bitmap and the last byte represents the pixels in the upper-right corner.

The biBitCount member of the BITMAPINFOHEADER structure determines the number of bits that define each pixel and the maximum number of colors in the bitmap. These members can take the any of the following values as depicted in the table given below:

| Value | Meaning |
| --- | --- |
| 1 | Bitmap is monochrome and the color table contains two entries. Each bit in the bitmap array represents a pixel. If the bit is clear, the pixel is displayed with the color of the first entry in the color table. If the bit is set, the pixel has the color of the second entry in the table. |
| 4 | Bitmap has a maximum of 16 colors. Each pixel in the bitmap is represented by a 4-bit index into the color table. For example, if the first byte in the bitmap is 0x1F, the byte represents two pixels. The first pixel contains the color in the second table entry, and the second pixel contains the color in the sixteenth table entry. |
| 8 | Bitmap has a maximum of 256 colors. Each pixel in the bitmap is represented by a 1-byte index into the color table. For example, if the first byte in the bitmap is 0x1F, the first pixel has the color of the thirty-second table entry. |
| 24 | Bitmap has a maximum of 2^24 colors. The bmiColors (or bmciColors) member is NULL, and each 3-byte sequence in the bitmap array represents the relative intensities of red, green, and blue, respectively, for a pixel. |

The biClrUsed member of the BITMAPINFOHEADER structure specifies the number of color indexes in the color table actually used by the bitmap. If the biClrUsed member is set to zero, the bitmap uses the maximum number of colors corresponding to the value of the biBitCount member.

The biCompression member tells whether the image is compressed or not. Windows versions 3.0 and later support run-length encoded (RLE) formats for compressing bitmaps that use 4 bits per pixel and 8 bits per pixel. Uncompressed images are taken for this project and they contain a value 0 in this field.

## 5. PROPOSED METHOD

The proposed method works with the original image as it is. It does not process the image in any way and transform the pixels of the image as in edge detection. It finds all the possible 4-connected contours and stores the 4-directions of the contour along with grey value being examined. The process is repeated with the help of a recursive procedure and marking all the pixels visited along the contour path. The marked pixels are eliminated for further examination of connected pixels[10]. The four direction crack code values (0,1,2,3 consuming 2 bits per number) are packed into a byte and stored along with the grey value in output file. No loss of pixels[1] are observed in the proposed compression method. The following is the format of

stored compressed image:

**Row, Column, Grey-Value, 4-direction crack codes**

### 5.1 Algorithm for Compressing Original Image

The following algorithm shows the sequence of steps to be followed to compress the original image.

**Step 1:** Read an uncompressed image file[6]

**Step 2:** Read number of rows n and columns m of the image from header

**Step 3:** :Separate pixels P[n,m]

**Step 4:** For i=1 to n do 5

**Step 5:** For j=1 to m do
   Store P[i,j] and its grey value g as beginning of the contour
   Mark the pixel P[i,j]
   **Crack_Code(P,i,j,g)**

**Step 6:** Write the header information and contour codes in another file.
**Procedure Crack_Code(P,i,j,g)**
Begin
   if (P[i, j-1] equal g) then store 0; Crack_Code(P,i, j-1,g);
      else if(P[i-1,j] equals g) then store 1; Crack_Code(P,i-1,,j,g);
      else if(P[i,,j+1] equals g) then store 2; Crack_Code(P,i, j+1,g);
      else if(P[i+1,,j] equals g) then store 3; Crack_Code(P,i+1, j,g);
      else return;
**End;**

### 5.2 Algorithm for Restoration of Original Image from Compresed Image

The following algorithm shows the sequence of steps to restore the original image from compressed image.

**Step 1:** Open the compressed image file.

**Step 2:** Read number of rows m and columns n of the image from header.

**Step 3:** Initialize P[n,m]

**Step 4:** Repeat steps 5 to 8 until all the crack coded contours are processed

**Step 5:** Read starting coordinate position(i, j) and grey value g of next contour.

**Step 6:** P[i, j]=g;

**Step 7:** Read next crack code c;

**Step 8:** Replace_Pixel(P,i, j,g,c);

**Step 9:** Write the header information and pixels P[n,m] in another file.

**Procedure Replace_Pixel(P,i, j,g,c)**
Begin
  if(c equals 0) then store P[i, j-1]=g;
    else if(c equals 1) then store P[i-1, j]=g;
    else if(c equals 2) then store P[i, j+1]=g;
    else if(c equals 3) then store P[i+1, j]=g;
    else  return;
**End;**

## 6. RESULTS AND DISCUSSION

The authors have developed a package using C language code for the proposed compression and decompression methods.  A set of sample bitmap images (both monochrome and color) are tested with the proposed method.  The compression percentage varies from 79% to 85% for the samples. The percentage of compression is better for images with more number of similar grey values[10]. No loss of pixels are found while restoring the original image.  Instead of storing 8 bits, the contour values are stored in 2 bits. Original size and compressed size of the images and computation time are plotted.

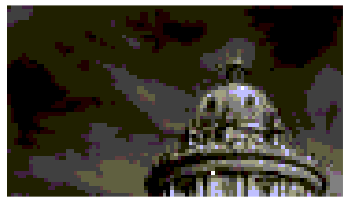

**Figure 4(a) Original Bitmap Image**

A sample content of the file which stores Starting Position of a pixel, Grey value and Crack Codes of its contour is shown below.  The last value -1 marks the end of the contour.
0 0 200 3 2 2 2 2 2 3 0 0 -1
0 2 225 1 3 3 2 2 2 2 2 2 2 2 3 0 0 0 0 0 0 0 0 3 2 2 2 2 2 2 2 2 3 0 -1
0 7 175 3 2 2 2 2 2 3 2 2 2 2 -1
0 9 180 2 2 2 2 -1
1 7 190 2 2 2 2 -1

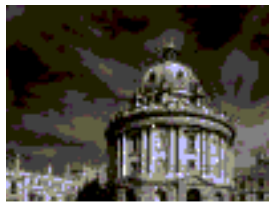

**Figure 4(b)  IMAGE after Decompression**

Table 3  Experimental Results

| SAMPLE BITMAP IMAGES | ORIGINAL SIZE IN BYTES | COMPRESSED IMAGE SIZE IN BYTES | COMPRESSION PERCENTAGE | COMPUTATION TIME IN SECONDS | NUMBER OF BYTES DIFFERED |
|---|---|---|---|---|---|
| 1 | 9108 | 1617 | 82.25 | 0.821 | 0 |
| 2 | 8036 | 1580 | 80.34 | 0.898 | 0 |
| 3 | 8415 | 1611 | 80.86 | 1.325 | 0 |
| 4 | 7698 | 1579 | 79.49 | 1.077 | 0 |
| 5 | 9078 | 1793 | 80.25 | 1.123 | 0 |
| 6 | 9783 | 1382 | 85.87 | 0.715 | 0 |

## 7. CONCLUSION

The proposed method proves to be a lossless compression method. Program execution time, compression percentage and rate of information loss is measured for various images. Computation time for compression of an image is not directly proportional to the size of the image.  It depends on the number of contours found in the image. As there is no loss of pixels, this method is more suitable for compressing medical images. The next phase of the research work with 8-connected pixels (chain coding) is under progress.

The International Journal of Multimedia & Its Applications (IJMA) Vol.3, No.4, November 2011

## AUTHORS

**Prof. Dr. T. Meyyappan** M.Sc., M.Phil., M.B.A., Ph.D., currently, Associate Professor, Department of Computer Science and Engineering, Alagappa University, Karaikudi, TamilNadu. He has obtained his Ph.D. in Computer Science in January 2011 and published a number of research papers in National and International journals and conferences. He has developed Software packages for Examination, Admission Processing and official Website of Alagappa University. His research areas include Operations Research, Digital Image Processing, Fault Tolerant computing, Network security and Data Mining.

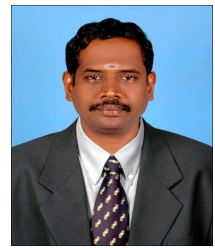

**Mrs S. M. Thamarai** received her Diploma in Electronics and Communication Engineering in Department of Technical Education,TamilNadu in 1989 and her B.C.A., M.Sc. (First Rank holder and Gold Medalist),M.Phil.(First Rank holder)degrees in Computer Science(1998-2005) from Alagappa University. She has published a number of research papers in International Journals, National and International Conference proceedings.Her current research interests are in Operational Research, Fault Tolerant Computing and Digital Image Proces sing. She is currently pursuing her Ph.D. in Alagappa University, Karaikudi, TamilNadu.

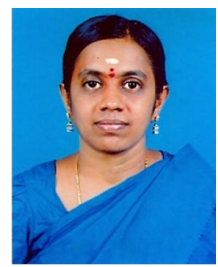

**Mr. N. M. Jeya Nachiaban** B.E. [pre-final] is pursuing Bachelor Degree in Computer Science and Engineering, Thiagarajar College of Engineering, Madurai. His research areas include Digital Image Processing and Data Mining.

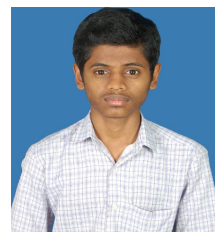